# PREFACE :

*Welcome,* all the delegates, scientists and students to the February 2007 seminar on POSITRON ANNIHILATION at **Saha Institute of Nuclear Physics,** Kolkata, India.

It is indeed a rare occasion that we find time to meet each other collectively for an objective discussion on a subject like 'Positron Annihilation' which encompasses various multidisciplinary areas in science.

I believe the message of the subject : that the smallest positive elementary particle, with all its potency is actually omnipresent and sees through all material architecture.

It actually depends on us 'the investigators' to decipher the tact of the nature and thus rejoice the understanding of the truth (as far as possible).

Thus, here we gather on a common platform to share each others view for a gallant discussion on different aspects. In our institute, we have had a long tradition of experimentation in Positron Annihilation Spectroscopy. Quite some time later during late eighties of the past century, another avenue Positronium Chemistry ( electron –positron bound state designated as lightest atom, serving as chemical probe) was started which subsequently has evolved to study chemical phenomenology ranging from liquid systems to various molecular substances and porous solids. This paved the way for physicists and chemists, to indulge in joint activities both in experimentation and theory to unravel the underlying truth behind the systems under investigation.

Currently we see in the literature, there have been phenomenal growth of the subject on polymers, micelles, nano particles, porous materials etc. We have tried to contribute in these areas which were reported in various symposia and journals. From our group we have also participated in international collaborations.

There are also new emerging areas and challenging task for Positron/Positronium scientists like possibilities of production of a large intensity of positronium atoms in a substance, possibilities of Bose-Einstein condensation of positronium atoms, improvements in molecular imaging technology, namely in PET, other biologically related activities, and surface (to depth of a few Angstroms) phenomena using mono energetic slow positron beams.

For the furtherance of research and investigation in the multi directional area, we also call up on the delegates from the universities around us who actually nurture the young scholars and handle teaching programme to participate in the seminar, so that this scientific area evolves gracefully in our country and we can work in a common platform with joint activities for the International Conference on Positron Annihilation, ICPA-15 to be hosted by Saha Institute of Nuclear physics.

*Bichitra Ganguly*
*Positron Annihilation Laboratory*
*Saha Institute of Nuclear Physics*
**February 9-10, 2006**

# The history of research with positrons
# at Saha Institute of Nuclear Physics, Kolkata

**P.M.G. Nambissan**

*Nuclear and Atomic Physics Division, Saha Institute of Nuclear Physics,*
*1/AF Bidhannagar, Kolkata 700 064*

(e-mail : pmg.nambissan@saha.ac.in)

Saha Institute of Nuclear Physics, Kolkata has a history of more than four decades of research using positron annihilation spectroscopy in materials of varying physical and technological interest. The universally realized potential of positron annihilation as a powerful probe for studying different types of solids and liquids was accepted way back in the early sixties of the last century by Prof. A.P. Patro, who had started a modest laboratory at that time exclusively for such studies. In the initial days, studies were focused basically on organic materials and solvents. Using the then available scintillation detectors and photomultiplier tubes of the best performance, a positron lifetime spectrometer was set up and a number of measurements were performed. Attempts were also made to measure the $3\gamma$ to $2\gamma$ ratio in some organic fluids using a Ge detector of good energy resolution. The papers were published in journals like Nuovo Cemento and had been cited by other researchers in related fields.

Prof. Prasanta Sen, who had done his Ph.D work under the supervision of Prof. Patro, rejoined the laboratory as a faculty member in the late seventies and it marked the second stage of resurgence of the activities in this field. He, along with his students, started the positron annihilation studies of defects in ion irradiated metals and semiconductors. The beams from the variable energy cyclotron available at the center next door opened up the possibilities of a wide range of research activities on the evolution of radiation-induced defects in solids under isochronal annealing treatments. Several scholars did dissertations for their Ph.D degree on the basis of the results from these experiments. The studies soon also focused on the dynamics of helium bubbles generated by uniform implantation and annealing. Several new information on the kinetics of defects and helium bubbles could be obtained from these studies.

In the late eighties and early nineties, a number of studies on materials of different aspects and interest such as quasicrystals, high temperature superconductors, geological samples and intermetallic compounds were carried out. Studies at low temperatures became routine with the procurement of a Leybold refrigerator cryostat.

Positron annihilation studies of nanomaterials were initiated in the laboratory as early as in 1995 and it continues even today as the most important activity in the laboratory. A large number of nanosystems of metals, semiconductors, ferrites and carbon have been already studied and a lot more are currently in progress. Many interesting aspects like structural and phase transformations, metal-to-semiconductor-like transition, quantum size effects etc. could be investigated and the results provided complementary experimental support to the findings from other more conventional but less sensitive techniques. Interest in ion-implanted solids has been recently revived with the availability of a variety of ion beams from the different accelerator facilities in the country.

Many scientists from different institutions within and outside the city of Kolkata have been doing collaborative experiments in the laboratory for the past several years. With experimental arrangements for positron lifetime, Doppler broadening and coincidence Doppler broadening spectroscopic measurements, the laboratory at present is capable of meeting the requirements of a large number of scientists engaged in research in different fields of solid state physics and material science. Although this brief survey has been widely qualitative in recapitulating the activities of eventful four decades, representative results available from the published work will be reported at the said meeting. References to the published works will also be made available.

------------------------------------

# Positron Annihilation studies of stainless steels and metal-semiconductor junctions

G. Amarendra*, R. Rajaraman, S. Abhaya, G. Venugopal Rao and C. S. Sundar
Materials Science Division,
Indira Gandhi Centre for Atomic Research,
Kalpakkam – 603 102, T.N, INDIA

Various positron annihilation studies carried out at IGCAR, Kalpakkam using fast positrons and variable slow positron beams will be highlighted in the present talk. Conventional positron lifetime and Doppler broadening measurements have been carried out on a variety of stainless steels. With regard to positron beams, Doppler broadening studies have been carried out as a function of positron beam energy, so as to obtain depth-resolved information in a variety of thin film structures.

Study of point defects, precipitates, their clustering and annealing in structural steel alloys is an important and fundamental area of research so as to obtain a comprehensive understanding of the material behavior under irradiation environment. Positron lifetime measurements on Ti-modified stainless steel(15% Ni, 14% Cr) D9 samples, used as a structural materials in prototype fast breeder reactor (PFBR), in solution annealed and 20 % cold-worked states have revealed clear stages attributable to point-defect annealing, nucleation and growth of titanium-carbide (TiC) precipitates. The influence of percentage cold work and the Ti/C ratio were also investigated. These results are further complemented with heavy ion Ni-irradiation studies with regard to identification of peak swelling temperature. The radiation damage occurring in intense neutron environment has been simulated using heavy ion Ni irradiation and the occurrence of voids in the near-surface regions of D9 sample have been probed by depth resolved S-parameter studies using positron beam.

Study of metal-silicides is an interesting area of research having technological applications in semiconductor industry. When a thin film of metal-silicon junction is subjected to elevated annealing temperatures, diffusion of either metal or semiconductor species across the junction takes place leading to intermixing and further, chemical reaction takes place between the metal and Si atoms resulting in the formation of metal-rich or Si-rich silicides. Detailed studies on Pd/Si, Ni/Si and Co/Si thin film junctions have been carried out using variable low energy positron beam, so as to elucidate the diffusion, defects and phase transformations in these junctions. The experimental results of positron beam are complemented with GIXRD, RBS and AES studies. In the case of Pd/Si, it is found that there is a production of excess vacancy-defects across the interface upon the formation of $Pd_2Si$ phase. For Ni/Si system, as the annealing temperature is increased it is found that different $Ni_2Si$, NiSi and $NiSi_2$ phases are formed and the final silicide phase $NiSi_2$ is devoid of vacancy defects. For the case of Co/Si, the influence of various synthesis methods viz., thermal evaporation, sputter-deposition and ion-implantation on the sequence of silicide phases has been investigated. It is found that the nature of the silicides phases formed is dependent on the initial microstructure of metal thin film in terms of its thickness, point defects and crystallinity. A comparison of these results will be presented. So as to investigate the sensitivity of annihilation characteristics to silicidation process, ab-intio calculations of positron lifetimes in various metal-silicides have also been carried out and these will be highlighted.

* Email: amar@igcar.gov.in

## Search for Field-assisted Moderators for Positron Beams

B. K. Panda

Institute of Physics

Bhubaneswar-751005

Orissa

Positrons implanted from a radioactive source into a material get thermalized by several interaction channels and then get emitted to vacuum provided the positron work function is negative. It is possible to apply an electric field to semiconductors with negative positron work function to drift thermalized positrons to vacuum for fabricating the field-assisted (FA) positron moderators in order to make a positron beam facility. This condition is satisfied when the positron affinity, which is the sum of positron and electron chemical potentials, is low.

Our method of calculating positron affinity in semiconductors is based on *ab initio* pseudo potential and linear-muffin-tin-orbital (LMTO) methods. While the electron chemical potential is calculated at the Fermi level with respect to the crystal zero, the positron chemical potential is calculated using the positron ground state energy from the same reference. The calculated positron affinities in pseudo potential and LMTO methods are compared with experiments in the following table with experiments in the following table.

| Material | Pseudopotential (eV) | LMTO (eV) | Experiments (eV) |
|---|---|---|---|
| Diamond | -2.24 | -2.42 | -1.15 |
| Si | -5.98 | -6.74 | -4.14 |
| SiC | -4.10 | -5.05 | -3.83$\pm$ 0.45 |
| BN | -3.37 | -3.94 | |

It is clear that the positron affinities calculated in the pseudopotential method is close to experiments with discrepancies arising from the experimental methods and defect states in semiconductors. Both diamond and BN for their low positron affinities are good candidates for fabricating FA positron moderators.

# Current problems and unknowns in positronium dynamics

**G. Duplâtre**
Nuclear Chemistry, IPHC
23 rue du Loess
BP 28 67037 Strasbourg
France
ULP/CNRS/IN2P3

**1. Introduction.** In spite of decades of studies on the chemistry of positronium (Ps), the bound-state of an electron with a positron, several questions regarding its behaviour as a chemical species have remained unanswered [1]. The importance of gaining understanding in this respect is twofold.

(i) On fundamental grounds, Ps may be considered as a paragon for chemical theoretical studies. As a normal chemical entity, it can, in principle, undergo most of the conventional chemical reactions; however, only a number of these has been firmly characterized. In addition, as a radical possessing a triplet (o-Ps) and a singlet (p-Ps) state, it may suffer spin conversion reactions. Some authors have also proposed that, due to its low mass, it would be liable to tunnelling effects [2].

(ii) As a probe of matter, the potential applications of Ps are manifold. Clearly, however, deriving quantitative information on a medium, whether liquid, solid or gaseous, requires a profound, as detailed as possible, knowledge of the Ps dynamics.

Although Ps chemistry has been long treated through very conventional approaches, two of its intrinsic properties make it an exceptional species. The first one relates to the possibility, for Ps, to build a bubble around itself when present in a liquid, due to the repulsive forces exerted on both the electron and the positron by the surrounding molecules. Several models have been proposed aiming essentially at defining the radius of the bubble formed; until now, all published expressions refer to the sole surface tension of the liquid as the driving parameter [3-5]. The existence of the bubble state should result in a completely different diffusion coefficient for Ps. The second property that makes Ps exceptional is its short lifetime which, in liquids, is most usually in the range 1 – 4 ns. Due to this, any reaction of Ps takes place over a time window where kinetic theories predict the reaction rate coefficient not to be constant [6]. Although such a constraint has been early known, and verified, for another light, transient chemical species, namely the solvated electron [7], it is only recently that it was recognized that it should apply also to the case of Ps [8]. The consequences of this time-dependence have not yet been fully valued, but it is clear that the span of its lifetime (ns) as compared to that of the solvated electron (μs) should make these consequences far from marginal as they are for the latter species.

Among the many problems remained unsolved regarding Ps chemistry, the present contribution will be focalized on the (hopefully, apparent) contradictions brought about when confronting experimental results with the two concepts just depicted: Ps bubble state (BS) and time-dependence (TD) of the reaction rate coefficient.

**2. General Ps chemistry**. Various examples of Ps chemical reactions, taken from the literature, will be scrutinized, with emphasis on oxidation reactions. The implications of BS and TD will be examined and discussed.

**3. Ps bound-state formation**. This particular category of reactions has been treated by many authors [1, 9 - 11], both experimentally and theoretically. However, quantitative approaches still show very important inconsistencies.

# Laser cooling of ortho-Positronium

**N. N. Mondal[a,b], T. Kumita[b], T. Hirose[b], H. Iijima[b], K. Kadoya[b], M. Irako[b], T. Matsumoto[b], K. Wada[b], and M. Kajita[c].**

[a] Saha Institute of Nuclear Physics, Kolkata, India.
[b] Tokyo Metropolitan University, Tokyo, Japan.
[c] University of Tokyo, Tokyo, Japan.
[a,b]e-mail: nagendra.mondal@saha.ac.in

## Abstract

Recently Bose-Einstein Condensation (BEC) is increasingly an interesting subject in atomic physics, nanomaterials as well as other branches of science. Positronium (Ps) is a bound state atom of electron ($e^-$) and its anti particle the positron ($e^+$), and that can be produced on a clean metal surface after bombardment of slow $e^+$ beam. Our interest is to achieve BEC in the case of Ps atom[1]. Laser cooling of ortho-Ps (o-Ps) is considered prior understanding of the PsBEC. Ps atom has extremely light mass (1/918 of H atom) and short lifetime. Thermal o-Ps (Spin triplet $^3S_1$ ground state, lifetime 142 ns) dominantly decay into 3γ rays, while para-Ps (p-Ps) a spin singlet ground state ($^1S_o$, lifetime 125 ps) dominantly decays into 2γ rays. Thermal o-Ps is the candidate of laser cooling system.

Theoretical studies of o-Ps laser cooling has been done by Liang et al.[2] utilizing 1S-2P transition, and that energy interval is 5.1 eV corresponding to the photon wavelength of 243 nm. Since the lifetime of the spontaneous transition from 2P to 1S is 3.2 ns, Doppler cooling of p-Ps is not possible for extremely short lifetime. A characteristic feature of the laser cooling of o-Ps is that the recoil energy is large owing to the small Ps mass, thus photon recoil limit (0.6 K) is higher than the Doppler limit (7.5 nK). In a sufficiently strong laser field, spontaneous emission from an o-Ps atom occurs every 6.4 ns on the average, which is two times longer than the lifetime of the 2P state because populations of 1S and 2P states are same. Thus the total cooling time is estimated to be 6.4×32 ≈ 200 ns, which is comparable to the lifetime of o-Ps. It can be noticed that the lifetime of o-Ps is also doubled in the strong laser field because of the direct decay time (100 μs) from the 2P state is much longer than that from the 1S state.

Wavelength of the cooling laser is usually swept during the cooling process to compensate Doppler shift due to motion of atoms while ordinary atoms are cooled. However this technique is not applicable for the o-Ps laser cooling because of the short cooling time. Thus a laser with wide line width, which covers Doppler broadening of the 1S-2P resonances without sweeping its wavelength, is required. The pulse duration also has to be low enough to cover the cooling time ≈ 200 ns.

In our extensive 3-Dimensionals Monte Carlo Simulation studies, the initial space and time distributions of the Ps cloud are set to 0.75 cm (1σ) on the x-y plane and 2.5 ns (1σ), corresponding to the size and pulse width of our $e^+$ beam respectively. Thermal o-Ps atoms obey Maxwell-Boltzmann distribution at 300° K. The pulse energy of cooling laser is assumed to be 60 μJ, time duration 68 ns (1σ), linewidth 11 pm (1σ) and radius 2.5 mm

---

(1σ). Laser beam is divided into six and irradiates the Ps cloud along the ±x, ±y, and ±z directions. As a result about 7% of generated o-Ps atoms are cooled down to 1K in 220 ns for laser detune of 170 GHz [3].

Positron beam is bunched with a laser pulse and injected on a target to produce Ps atoms. The width of the pulse is 15 ns (FWHM) with the intensity of 0.005 e$^+$s/pulse while it is operated with 100 kHz repetitions. Cr:LiSAF is chosen because of its long lifetime of the excited state (67 μs) and low gain, which results long pulse duration [4]. In order to detect laser cooled o-Ps atoms a novel detection system has been developed [5]. Advancement towards Ps-BEC will be discussed.

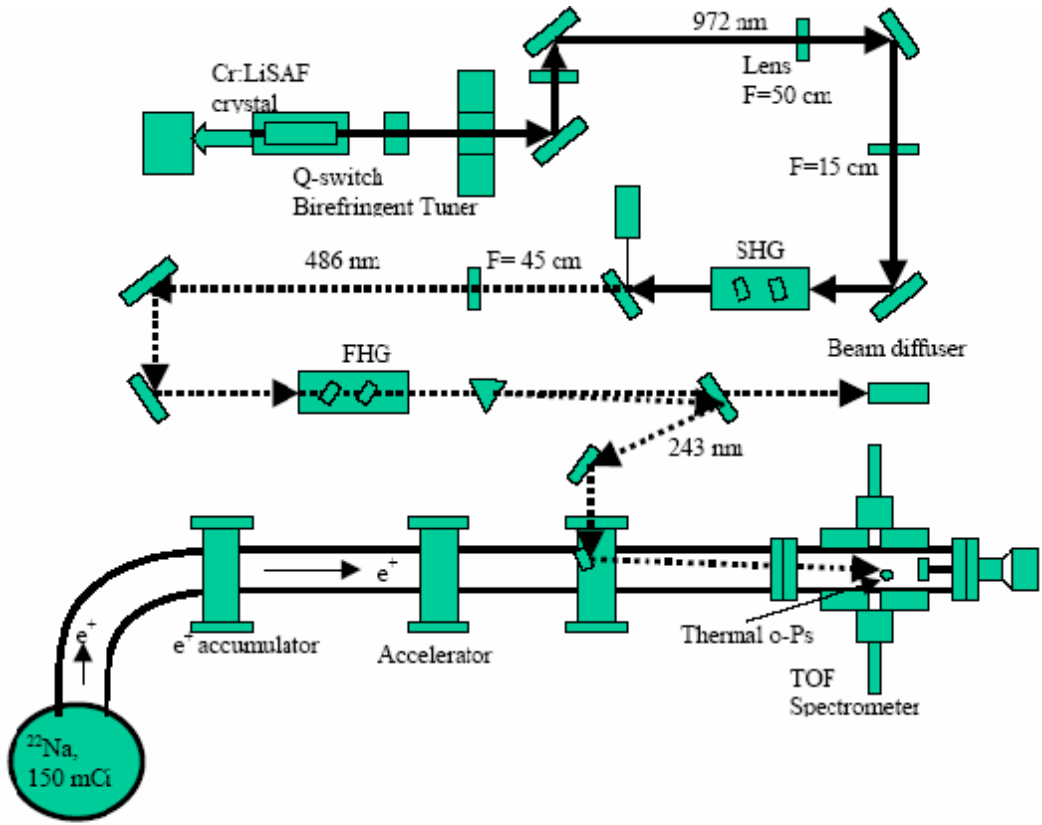

A schematic view of the laser cooling system with e$^+$ beam and TOF system

# Positron annihilation studies in porous media

D. Dutta, K. Sudarshan, S. K. Sharma and P. K. Pujari

Radio Chemistry Division, Bhabha Atomic Research Centre, Trombay, Mumbai-400 085

Applicability of positron annihilation technique as an in situ probe for study of porous material is well documented. Despite a lot of work reported in literature, there is considerable scope to understand the positron/Ps systematics in porous solids. In this presentation we would highlight our studies where in a dependence of $I_3$ & $\tau_3$ on the sampling fraction of 3-gamma component have been demonstrated. The intensity of Ps is known to increase with sampling of 3-gamma fraction. We have, for the first time, observed the increase in $\tau_3$ which has been explained on the basis of pore size distribution.

In addition, we have studied phase transition of liquids confined in nanopores of ZSM-5 zeolite and silica gel material using temperature dependent Doppler broadening and lifetime spectroscopy. The pore diameters were in the range of 7-60 Å. Benzene confined in intergranular spaces (meso-pores, pore diameter $\geq$ 20 Å) and micropores (pore diameter < 20 Å) of the porous material shows distinct freezing points. While the freezing temperature of benzene in mesopore is close to the bulk value (278K), a large depression is observed for benzene inside micropores. It is observed that the freezing point decreases with decreasing pore size. A linear correlation is obtained between the depression of freezing/melting point and the reciprocal of the pore diameter, which does not exactly follow classical Gibbs-Thomson relation.

The reactivity of nitrobenzene, a known quencher of positronium, has also been studied in nitrobenzene-benzene mixture incorporated in ZSM-5 pores. An anomalous enhancement in the reactivity is observed as compared to measurements in the bulk liquid. On the other hand, various amounts of nitrobenzene absorbed on the surface of pores of XAD-4 resin and show a continuous decrease in the ortho-positronium intensity with increasing nitrobenzene loading. No appreciable change in the ortho-positronium lifetime is observed. This suggests that the nitrobenzene acts as inhibitor rather than quencher in XAD-4 resin. The results and possible reasons behind these anomalous behaviors will be discussed.

# A Theoretical Investigation of Positron Molecule Collision at Low Energies

**Tapas Mukherjee**

Physics Department, Bhairab Ganguly College,
Kolkata- 700 056
India

## Abstract

The amazing prediction of Dirac and subsequent experimental confirmation by Anderson and Blackett and Occhialine for the existence of positron pioneered the era of positron physics. One of the major fields where the positron is employed as a probe is molecular collision physics. The analysis of collision phenomena quantum mechanically provides the detail laws of interactions as well as the structure of matter on microscopic scale. Here we study the positron molecule collision using Eigen State expansion or close-coupling method where the total wave function of the system is expanded in terms of the Eigen states of the molecules. However, the appearance of nuclear degrees of freedom, multi center nature of positron molecule interaction, non spherical nature of potential and existence of permanent dipole for polar molecule make the problems more complicated than atomic collision. So one has to invoke the approximation, so called Born-Oppenheimer (B-O) approximation where the slow variables (nuclear coordinates) and the fast variables (electronic coordinates) are separated, so that one can first solve the electronic problem, with the nuclei fixed, obtaining electronic wave functions and energies that depend parametrically on the nuclear coordinate. However, in solving the scattering problem under B-O approximation one tries to restrict the nuclear degrees of freedom to make the calculation tractable. In the calculations, approximating the total Hamiltonian of the system, which leads to different models being employed, imposes the restriction of nuclear motion. The models are (1) Fixed Nuclei approximation (FN), (2) Adiabatic Nuclei approximation (ANR, ANV), (3) Rigid Rotor Model (LFCCA), (4) Ro-vibrational approximation (VIB-LFCCA), (5) Body Frame Vibrational close coupling (BFVCC), (6) BFVCC-Adiabatic Angular Momentum Coupling (BFVCC-AAMC) and most general (7) Ro-vibronic method. Here we have discussed some of the methods used to calculate scattering parameters at low and medium energies. Main motivation of this presentation is to compare and to find out the dynamical coupling effects between different degrees of freedom. For this we have presented some of the results obtained by us for positron molecule collisions, the molecules being hydrogen, carbon monoxide, hydrogen molecular ion, oxygen, nitrogen, carbon-dioxide etc. The results demonstrate the necessity of the use of more rigorous close coupling methods mainly for the cases, (1) Near the rotational excitation threshold, (2) Near the electronic excitation threshold, (3) For the system where the long range force dominates, (4) For the system where the molecule is not in the rotationally zero state, (5) In the energy region where the effect of vibrational motion of the nuclei is appreciable. Moreover, we have seen that the coupling between the positron angular momentum and that of the target can not be simplified as it can occur in the case of weaker vander Wall's interaction for which such a method viz. Adiabatic Angular Momentum coupling (AAMC) is usually suggested.

**Positron annihilation studies on pervaporation membranes**

**Dr. V S Subramanium**
*Benaras Hindu University*
**Email** : vsspt@yahoo.com

Dense membrane pervaporation is a separation technology of current interest in separating liquid mixtures, particularly azeotropic mixtures. This process requires the use of dense membranes that are non-porous. The separation is carried out using dense polymeric films. The mixture of the liquid to be separated is fed on one side of the membrane in a cell and the permeating flux, in the vapour form, is obtained on the other side of the membrane by applying a pressure lower than the saturated vapour pressure. The free volume in the membrane plays an important role in the pervaporation process. Positron annihilation techniques, especially the lifetime technique has developed into a powerful characterization tool for the study of free volume holes in polymers. In this work, we attempted to understand the relation between the free volume hole sizes, obtained from positron annihilation lifetime spectroscopy, of some commercial and laboratory made pervaporation membranes, and their pervaporation performance.

We also studied the effect of dibutyl phthalate plasticization (0 to 6 wt%) of ethyl cellulose polymer using positron annihilation technique. Free volume hole content has been found to increase as a function of plastcizer addition, with a mild tendency to saturate at higher concentrations. The intensity of the long lifetime component is almost a constant in the entire concentration range, indicating that the free volume hole content change is entirely due to the change in the size of the free volume holes. The results emphasize the applicability of positron annihilation techniques for the characterization of free volumes in polymer membranes used in liquid mixture separation technologies.

**POSITRON LIFETIME SPECTROSCOPY AND LIQUID CRYSTALLINE MATERIALS**

R. Yadav[1], K. Chandramani Singh[2], P. H. Khani[3], P.C. Jain[4] and A.N. Maitra[5].

[1] Department of physics, R.K. College Madhubani (Bihar) -847211
[2] Department of physics, Sri Venketesra College, University Of Delhi, New Delhi
[3] Department of physics, Faculty of Science, IHU, Teheran, Iran
[4] Department of physics and astrophysics, University Of Delhi, New Delhi.
[5] Department of chemistry, University Of Delhi, New Delhi

The main focus of our research work is to locate the various phases, structural and micro environmental changes occurring in liquid crystalline materials by employing positron lifetime spectroscopy. In some materials this approach has resulted in detecting some new phases not revealed by other conventional techniques. Another approach in these investigations has been to understand the mechanism of positron annihilation in these materials. In thermotropic liquid crystals, the positron annihilation parameters are in general found to be very sensitive to the phase transition. The nature and the magnitude of the changes depend on the type of the transition and the materials. In lyotropic liquid crystalline systems, changes were observed in the positron lifetime parameters whenever a phase transformation occurred. The various phase boundaries demarcated by this technique agrees well with those obtained by other conventional techniques. Besides this, the technique provides finer details of an otherwise considered being a single phase region. The existence of new structures has been demonstrated by a change in the trend of o-Ps lifetime. However, the positron annihilation mechanism in molecular materials is still far from being fully understood, the results obtained from this technique are still qualitative in nature. To quantify them, more systematic work is required to provide a better understanding of the positron annihilation mechanism in such systems

# Current areas of research using positrons in materials at Saha Institute of Nuclear Physics

**P.M.G. Nambissan**

*Nuclear and Atomic Physics Division, Saha Institute of Nuclear Physics, 1/AF Bidhannagar, Kolkata 700 064*

(e-mail : pmg.nambissan@saha.ac.in)

## Abstract

A brief review is presented highlighting the use of positron annihilation spectroscopic techniques for exploring the interesting structural modifications accompanying the reduction of solids to composites of nanometre-sized particles. Even though a direct correlation of the changed annihilation characteristics with predictable changes in the solid is certainly not possible, positron annihilation can be used for obtaining complementary evidences to substantiate such predictions. Different types of materials exhibiting different types of transitions have been investigated and the results powerfully portray the sensitivity of positrons to changes in the electron density and momentum distributions in the annihilation sites. Measurements of positron lifetimes and the lineshape parameters of the Doppler broadened gamma ray spectra are carried out as functions of either mean grain sizes or temperature and from the variations of these quantities, defect-specific information are extracted, which are then interpreted in terms of the atomic rearrangement within the solid.

# Introduction

The principles of positron annihilation in condensed materials can be successfully utilized to explore several interesting aspects of materials. In particular, those materials composed of grains of nanometer dimensions brought to physics a whole lot of excitations from the structural as well as application points of view. Over the years, the studies of nanoparticles and systems composed of them have become subjects of in-depth studies in several branches of science like physics, chemistry, metallurgy and engineering. A number of processes and properties of nanophase materials depend on the atomic arrangement on the grain interfaces. Hence experimental probes such as positron annihilation will be very useful in understanding the properties of such interfaces. The results of some recent investigations are presented in this paper. The results presented are rather brief since the details have been already published in the papers indicated by appropriate references. The reader is also advised to familiarize himself/herself with the fundamental principles of positron annihilation, and how information is derived from experimental observations, by reading at least a few of the popular review articles on the subject [1-3].

## Experimental details

The details of the sample preparation are available in the respective papers cited as references. Samples prepared through different methods have been used, depending upon the particular sample of interest. Elemental nanocrystals like Nb and Ti, for example, were obtained through high energy ball milling for several hours. The nanocrystalline ferrites $ZnFe_2O_4$ and $NiFe_2O_4$ were prepared through the normal sol-gel method. Chemical routes have the advantage that the particle growth can be efficiently controlled through careful selection of the reagents and experimental conditions in order to obtain a very narrow distribution of the nanoparticles. Appropriate modifications are made to obtain nanosystems of different geometry such as ribbons, rods, wires and tubes.

The lifetimes of positrons of $^{22}$Na source subsequently annihilating in the sample under study are measured using a standard gamma-gamma ray coincidence setup of prompt time resolution 200 ps (and later improved to 160 ps) for gamma rays from a $^{60}$Co source. The positron lifetime spectra were analysed using the programs RESOLUTION and POSITRONFIT [4]. Doppler broadening measurements were carried out by using an HPGe detector of good resolution (1.2 keV at 511 keV) and a spectroscopy amplifier with good gain stability for long data acquisition intervals. A pulse shaping time 2 $\mu$s has been used while processing the signals. Coincidence Doppler broadening (CDBS) measurements [2] were done using one more HPGe detector of identical resolution in anticollinear geometry with the first one and generating a two-parameter coincidence spectrum, which was analyzed using the standard procedure [5,6].

## Results and discussion

For a proper understanding of the results and their interpretation, let us start with an elemental nanoparticle system. High purity Ti was ball-milled for several hours and samples were collected after various milling hours. Earlier x-ray diffraction studies had indicated a transformation of the hcp structure to fcc while reducing the grain size. A similar case was found in the case of bcc Nb, which transformed to fcc at grain sizes below about 10 nm [7]. Positron lifetimes showed characteristic changes highlighting such a changes in the system. Fig. 1 shows how the positron lifetimes and the relative intensity of one of them varied during the ball-milling of Ti [8]. Note that the positron lifetime predicted by the two-state trapping model [9], shown by the shaded circles in the figure, did not agree with the experimental results. The trapping of positrons in nanocrystalline materials does not strictly follow the mechanism underlying the aforesaid model. Quite often, even the shortest lifetime component $\tau_1$ could indicate the trapping of positrons in a defect at the surface of the grains. A typical case is that of the nanoparticles of zinc ferrite ($ZnFe_2O_4$), which, when investigated by positrons, revealed a systematic variation of their lifetimes and changes in the lattice constants and thereby the size of the tetrahedral and octahedral vacancy clusters [10]. Here positron lifetimes increased while reducing the grain size and finally decreased owing to the trapping in octahedral vacancies. Recently, the inverse spinel ferrite $NiFe_2O_4$ was studied by measuring positron lifetimes in samples of varying grain sizes. In this case, the nickel ferrite nanoparticles were synthesized within an amorphous $SiO_2$ matrix and, through proper model-based analysis, the effect due to decreasing particle size could be demonstrated [11]. Fig. 2 illustrates the lifetimes derived for those positrons annihilating within the nanoparticles while the particle size is gradually reduced. A remarkable increase at very low particle sizes has been interpreted as due to an inversion of the inverse spinel structure to the normal phase and was substantiated through Mossbauer spectroscopic studies.

An almost identical transformation is that of maghemite ($\gamma$-$Fe_2O_3$) nanoparticles to hematite ($\alpha$-$Fe_2O_3$), which realistically is temperature-induced rather than a finite size effect. Nevertheless, the temperature necessary for inducing the transformation has been significantly reduced upon nanocrystallisation of the material and hence it is an indirect manifestation of the structural instability caused by having only finite number of mutually adhering molecules. The measured positron lifetimes indeed reflected these changes, albeit at temperatures earlier than that indicated by x-ray diffraction patterns. The change in positron lifetime was also correlated with the change in volume of the respective unit cells of the two phases [12]. It was remarkable that the shorter lifetime $\tau_1$ reflected the characteristic decrease conspicuously owing to its vacancy-specific annihilation within the grains and thereby sensing the atomic redistribution around (Fig. 3). $\tau_2$ results from the annihilation at the intergranular region and decreased owing to the shrinkage of this region due to neighbouring grains increasing in size [12].

The attribution of the various positron lifetime components to the various trapping sites in nanomaterials is not exactly straightforward and is highly sample-specific. In the case of elemental nanoparticles where the interior of the grains are presumed to have the same perfect crystalline structure as that of the macrocrystalline counterparts, positron trapping within the grains can be practically ruled out and all the positrons can be assumed to have annihilated at the grain interfaces. This may not be the case with complex solids with uncommon structure like spinels or zinc-blend. For example, in $ZnFe_2O_4$ and $NiFe_2O_4$ nanoparticles, the inversion of the spinel structure could be sensed through positron lifetimes only since the positrons got trapped into the vacancies at the octahedral sites within the grains. Indeed structural transformation will modify the grain in totality, although the effect of the changes on the surface layer of atoms may not be reflected as strongly as within the grains, due to the highly diffusive nature of the atoms or vacancies on the grain surfaces.

In an experiment conducted to explore the sensitivity of positrons to nanosystems of varying geometrical forms and shapes, we observed that the positron lifetimes were characteristically different for different samples of the same material [13]. We recorded the positron lifetime spectra of $FeS_2$ samples of different nanometrical designs – grains, ribbons, rods, wires and tubes. A diagrammatic representation of the mean positron lifetimes associated with the various $FeS_2$ nanosyatems is illustrated in Fig. 4.

Several other nanosystems, both of metals and semiconductors, and including even other ferrite materials have been studied at different times in our laboratory at Saha Institute of Nuclear Physics and the details as well as interesting findings of those experiments like the one related to quantum confinement effects etc. can be found in the various publications [14-18]. Very recently, we have also initiated some works on polyacrylonitrile-based carbon fibres embedded with multi-wall carbon nanotubes [19]. The findings of these studies have been strongly and adequately supported by results of other complementary techniques like scanning electron microscopy and Raman spectroscopy. Currently we are studying ZnO nanocones (pure and Mn-dopped) and nanoparticles and the results will be soon reported elsewhere.

The interest in ion implanted solids was renewed recently with an investigation into the evolution under isochronal annealing of boron-induced defects in dopped Si. Fig. 5 illustrates the variation of the main defect component lifetime $\tau_2$ with the annealing temperature. The different stages of annealing and the supportive arguments from CDBS experiments are recently reported [20]. Recently we had carried out an experiment on the defects generated by boron irradiation of Fe-Al alloy and their annealing behaviour under different heat treatments had been monitored. The data analysis is in progress.

## Conclusions

The results presented in this review article clearly demonstrate the ability of positron annihilation techniques to explore several interesting aspects of materials including nanoparticles and related systems. Positron lifetimes and the lineshape parameters of the Doppler broadened annihilation gamma ray spectra are markedly different in nanosystems from their coarse-grained bulk counterparts. Normally, such changes can be attributed to the increase in the net defect volume at the increasing number of grain interfaces in nanomaterials. However, we also observed that a number of phase transitions also accompanied the nanocrystallisation of solids. Although the exact reasons for such transitions are topics of separate investigations, the structural instability both within and on the surface of the nanoparticles could possibly be driving such structural changes. For example, the bcc to fcc transition in nanocrystalline Nb also indicated a concomitant lattice expansion [7]. On the other hand, we observe the inversion of the spinel structure of $ZnFe_2O_4$ at very small particle sizes where the lattice got contracted [10]. The inversion of the inverse spinel structure of $NiFe_2O_4$ also occurred after an abrupt and substantial contraction of the lattice at very small grain sizes [11]. In all these cases, there has been a characteristic change in the positron annihilation characteristics on either side of the transition and hence the technique can be successfully used to get complementary evidences for such abnormal changes in nanophase materials. At the same time, the author wishes to emphasize that conventional areas of defect spectroscopy such as irradiated solids and inert gas bubbles in solids can still offer a fruitful avenue of investigation using positrons for quantitative analysis of their causes and effects.

## *Acknowledgments*

The author is grateful to the large number of collaborators who have contributed to the success of the different experimental works presented in this paper.

# POSITRONIUM STUDIES IN CHEMICAL ASPECTS

**Bichitra Nandi Ganguly**
**Saha Institute of Nuclear Physics**
**1/AF Bidhan Nagar, KOLKATA-700064.**
email :bichitra.ganguly@saha.ac.in

The bound state of electron and positron exists as a quasi stationary state known as positronium atom (Ps) which is designated as the lightest analogue of hydrogen atom (though it also differs in many ways) is being used in diverse fields as a chemical probe. Using the positronium interaction in the matter, one can basically understand the electronic organization of the same, mainly by two simple techniques: namely, positronium life time spectroscopy and Doppler broadening spectroscopy of annihilation radiation (or angular correlation spectroscopy). While the former technique yields the information about the electron density (or the change in electronic environment), the latter would infer on the energy state of the electrons. Thus, the annihilation characteristics of $(e^+e^-)$ contains all the information of the annihilation site and studying chemistry with this powerful probe opens up a fascinating area.

In the beginning, we had started studying the aggregation behaviour of the surfactant molecules as the positronium annihilation characteristics would sense the changes in electronic micro environment and this paved the path for investigation in micellar , reversed micellar and micro-emulsion systems[1]. Further the solubilization process of molecules or ions, their micro environment and the binding constant of these species in the substrate medium was determined through the kinetic rate constants of positronium-molecule interaction. This in fact has a correlation with the biological systems. We had also attempted to study the phenomenon of hydrotropism expressed by the aqueous solution of alkali metal salts of hydroxyl aromatic acid, such as salicylic acid.[2] The enhanced power of solubilizing sparingly soluble substances that were otherwise insoluble in water, is a special feature of this class of compounds. This is referred to as the hydrotropism and its importance in medicine is well known.

We have also studied the fundamental and intriguing features of Positronium complex formation with a conventional electron acceptor molecule like nitrobenzene(i.e. [Ps-M] bound state formation), in terms of temperature and pressure changes including its solvent dependence[3]. The mechanistic aspect of this study has invoked positronium bubble formation in the inert molecular liquids which were used as the reaction media[4]. It is indeed a very fascinating to follow how the Ps- bubble would influence the chemical reaction rate constants with temperature and pressure.

We have also ventured on the study of polymers. The different types of study include conducting polymers[5], radiation damage of polymers[6], porous membrane materials[7] etc. We have also studied porous materials like silica gel , zeolites etc. which are important in technology, demonstrated the usefulness of Ps as a probe to study the pore size both for the micro as well as meso porous substances[8]. Ps interaction in the pore surface constitute and its temperature dependence[9] involves an in depth introspection and further investigation.

**Acknowledgement :**

This abstract is the representation of the work that has been pursued over the decade by the Positronium study group, by different collaborators from time to time and my students who have contributed through their untiring efforts.

# Positron annihilation spectroscopic studies on some oxide materials

**Sreetama Dutta[1], Mahuya Chakraborty[2], S. Chattopadhyay[3], A. Sarkar[4], D. Sanyal[2] and D. Jana[1]**

[1]*Department of Physics, University of Calcutta, 92 Acharya Prafulla Chandra Road, Kolkata 700 009, India*
[2]*Variable Energy Cyclotron Centre, 1/AF, Bidhannagar, Kolkata 700 064, India*
[3]*Department of Physics, Taki Goverment College, Taki 743429, India.*
[4]*Department of Physics, Bangabasi Morning College, 19 Rajkumar Chakraborty Sarani, Kolkata 700009, India*

Positron annihilation spectroscopy has long been known as a simple elegant non-destructive technique for characterizing defects in solid materials [1,2]. Since the nature and the abundance of these defect centers control the transport and other physical properties of a system, the utility of this technique has been remarkably increased in the study of material science. We have successfully used positron annihilation lifetime (PAL) as well as coincidence Doppler broadened positron annihilation γ-radiation line-shape (CDBPARL) techniques for proper characterization of defects in various systems such as HTSCs, II-VI semiconductors, rare-earth manganites and other oxide materials [3-13]. Besides characterizing defects in different types of materials we have also worked in the development of experimental set up for both PAL and CDBPARL measurements. As a mater of fact, we can reach a time resolution of 182 ps and peak to background ratio of 14,000: 1 [4] during PAL and CDBPARL measurements respectively. All of these will be discussed in details.

# Applications of Positron Annihilation Spectroscopy to Polymeric and Biological Systems

**Y.C. Jean[1,2,3] , Hongmin Chen[1], Guang Liu[1], Lakshmi Chakka[1], Joseph E. Gadzia[4]**

[1]Department of Chemistry, University of Missouri Kansas City, Kansas City, MO 64110, USA
[2]R&D Center for Membrane Technology, Chung Yuan Christian University, Chung-Li, 32023, Taiwan
[3]Department of Chemical Engineering, Chung Yuan Christian University, Chung-Li, 32023, Taiwan
[4]Dermatology, Department of Internal Medicine, University of Kansas Medical Center, Kansas City, KS 66103 and Kansas Medical Clinic, Topeka, KS 66614

## Abstract

*Positron annihilation spectroscopy (PAS) is a novel radio-analytical technique which uses the positron (anti-electron) and is capable of probing the atomic and molecular scale (0.2-2 nm) free-volume and hole properties in polymeric and biological materials. Recently, we developed positron annihilation lifetime and Doppler broadening of energy spectroscopies coupled with a variable mono-energetic positron beam to measure the free-volume depth profile from the surface, interfaces, and to the bulk. This paper presents applications of PAS to determine multi-layer structures, glass transition temperatures in nano-scale polymeric films and to detect cancer in the human skin.*

## Introduction

The positron is the anti-electron. When the positron encounters electrons, it annihilates into γ-rays by Einstein's equation $E=mc^2$. The characteristics of annihilation photons contain electronic properties, i.e. wave function, density, and energy levels of matter at the location where the positron annihilates [1,2]. Positron annihilation spectroscopy (PAS) is a branch of γ-ray spectroscopy which monitors the lifetime, energy spectrum, and angular correlation of annihilation photons. PAS has been used to study defects in solids for many decades [1]. Recently, it has been successfully applied to chemical systems [2] to measure the free volume properties in polymers and in biological systems [3].

With its unique sensitivity to the atomic-level free volume in polymers, PAS is emerging as a promising tool to measure free-volume properties as a function of depth when one employs a variable mono-energy positron beam [4-6]. Positron annihilation lifetime (PAL) spectroscopy [2] is capable of determining size, quantity, distribution, and relative fraction of free volume in polymers due to the fact that the ortho-Positronium (o-Ps, the triplet Positronium) is preferentially trapped in the subnanoscale free volume. Similarly, Doppler broadening energy spectroscopy (DBES) [2] is another efficient technique to probe para-Positronium (p-Ps, singlet Positronium) and chemical composition of chemical and biological systems. In this paper, we present two applications of PAS: the nanoscale multi-layer structures and $T_g$ depth dependence [7] in ultra-thin polystyrene films supported on the Si substrate and the selectivity of positron annihilation with skin with and without cancerous cells [8,9].

## Experiments

The polystyrene (PS) used in this study was purchased from Aldrich Chemicals ($M_w$ = 212,400, $M_w/M_n$ = 1.06). Silicon wafers were from Wafer World Inc. (West Palm Beach, FL). The polystyrene films were prepared by dissolving different wt % polystyrene into toluene, then spin-coated onto Si wafers at a spin rate of 2000 RPM for 5 min. The films were annealed in vacuum at 140 °C for 12 h before mounting on a Kapton film heater in a vacuum system for PAS measurement. The film thickness was found to be 82 ± 5 nm on Si using profilometry (Tencor alfa-step 200, Adv. Surf. Tech., Cleveland, OH) as described elsewhere [7]. The surface roughness of Si wafers was measured to be 3 ± 1 nm.

Cancerous samples containing 10%-50 % Basal Cell Carcinoma (BCC) and Squamous Cell Carcinoma (SCC) were surgically removed from living human skin as part of the Mohs Microsurgical technique [8], which entails resection of the tumor and horizontal frozen tissue processing for visualization of the tumor margins. The specimens used in our experiment however were not subjected to Hematoxylin and Eosin staining so that there is no interference due to the presence of dye. The sections used had an approximate diameter of 1-cm and were 15-μm thick. The specimens were placed on a glass plate and transported in dry ice from the medical facility to our positron annihilation laboratory for PAS experiments using a variable energy positron beam [7,8].

The positron beam (collimated to 0.5-cm diameter) under a vacuum of $10^{-5}$ Pa has a variable energy from 100 eV to 30 keV which corresponds to a mean depth from 1 nm to 8 μm according to an established equation [4]. Each PAL and DBES spectrum contains a total of one million counts. The PAL lifetime resolution was 250 ns for the conventional set-up and was 450 ps at the positron energy > 1 keV, respectively and is larger at lower energy and the energy resolution of solid state detector in DBES was 1.5 keV at 511 keV. Detailed descriptions of PAL and DBES using our positron beam can be found elsewhere [5,6,10]. A major part of the PAL data (< 25 ns) was fitted into three lifetimes using the PATFIT program and with the long lifetime ($\tau_3$) being assigned to o-Ps with intensity $I_3$. $\tau_3$ is used to calculate the mean radius of free volume and along with $I_3$, to indicate the relative fractional free volume in the samples to an established method [2].

## Results and Discussions

### 1. Layer Structures of Nano-Scale Films

Figure 1 shows the variation of the S parameter measured by DBES as a function of positron energy or mean depth (top) as calculated from the positron energy according to an established equation [2] for different thicknesses of PS prepared with different wt % of PS (0.3%, 0.5%, 0.8%, 1.0%, 1.5%, and 2.0%) in toluene on Si and spun dry [7]. A low value of S parameter at low positron energy for thin films is due to back-diffusion of the implanted positron and Ps from the polymer.

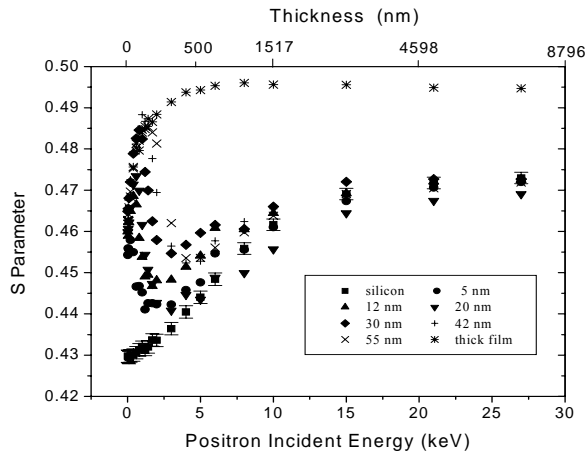

**Fig. 1** Plot of S parameter vs. implantation positron energy (depth) for blank silicon wafer, polystyrene films in different thicknesses, and a thick polymeric film [7].

When the positron reaches the interfacial region of polymer and substrate, the S parameter starts to decrease again as shown in Figure 1. A thinner film shows that the peak is closer to the surface. We define the full-width-half-maximum (FWHM) of the peak as the thickness of polymer film. They are found to be 5 nm (0.3%), 12 nm (0.5%), 20 nm (0.8%), 30 nm (1.0%), 42 nm (1.5%), and 55 nm (2.0%) with wt % of polystyrene in toluene as shown in parenthesis.

Figure 2 shows the variation of S parameter with respect to the positron incident energy and the mean depth (top x-axis) as calculated from an established equation for an $80 \pm 8$ nm thin polystyrene film (prepared from 3.0 wt % of PS in toluene) on Si wafers. From Figure 2, four regions for a polystyrene film are identified: the near-surface (region I), the polymeric film (region II), and the interfacial layer (region III) between the polymer and the Si substrate, and the bulk of the substrate (region IV). Although it is known that the depth of the implanted positron spreads out as the positron enters further into the bulk, the division of three regions in a thin polymer film is still clearly identifiable from S parameter vs. energy plot, as shown in Figure 2. We fitted the S variation with the depth in a multi-layer model by using a computer program VEPFIT [11], which has taken the positron implantation profile into consideration. The resolved film thickness ($80 \pm 8$ nm) from VEPTFIT analysis agrees well with the result obtained from profilometry ($82 \pm 5$ nm) [7]..

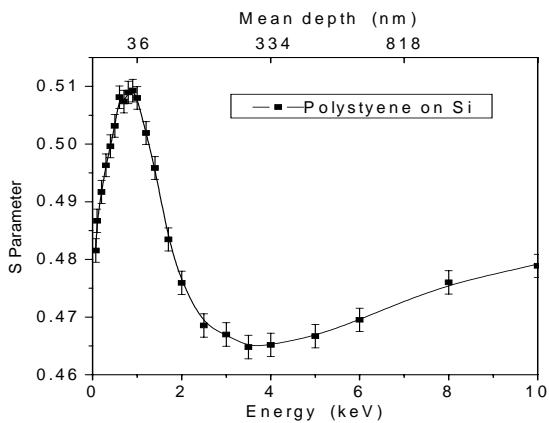

**Fig. 2** Plot of S parameter vs. positron incident energy (depth) in an 80-nm PS film (top) and depth (y-axis) [7]. Line was fitted in a multi-layer structure.

Figure 3 is a schematic diagram which shows the result of fitted nanoscale layer structure for a thin polymer film on substrate: surface layer (I), 80 nm film layer (II), interface layer (III), and Si substrate layer (IV).

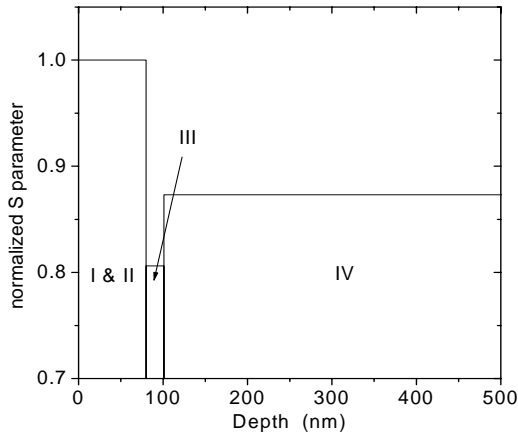

**Fig. 3**. A sketch of the layer structure of a polymer thin film on Si substrate: near surface of polymer film (I), the film (II), the interface between polymer film (III), and substrate (IV) [7].

The main conclusions from the fitted VEPFIT results of the S parameter vs. depth for an 80 nm PS film are: the density of the interface layer ($0.4 \pm 0.3$ g/cm$^3$) is significantly lower than that of the film layer (1.1 g/cm$^3$); the thickness of the film layer and the interface layer is $80 \pm 8$ nm and $21 \pm 3$ nm, respectively; the Ps diffusion length is very short, $3 \pm 1$ nm, and the positron diffusion length in the film and in the interface are $40 \pm 5$ nm and $9 \pm 3$ nm, respectively. The S parameter data are thus useful for the determination of layer structures for polymeric thin films.

Figure 4 shows the o-Ps lifetime $\tau_3$ and intensity $I_3$ vs. energy or mean depth in an 80 nm Ps film. The o-Ps lifetime is large at the surface and decreases to 2.1 ns at the film layer, then further decreases after the interface layer to 1.7 ns in the substrate. The observed o-Ps lifetime (1.7 ns) component in Si at a very small intensity ($I_3 < 0.5\%$) is attributed to some Ps annihilation with different depths of the film as the positron energy is spread out when the positron enters the substrate. It is interesting to observe that the variation of o-Ps intensity ($I_3$ of Figure 4) is similar to that of the S parameter in Figure 2. This is expected because S mainly determines p-Ps, while $I_3$ is the intensity of o-Ps. At the surface, the low intensity is due to back-scattering of positron and Positronium. Similar to S data, it increases to its peak value at the film layer, then decreases at the interface layer, then to nearly zero due to no o-Ps formation in the Si substrate.

## 2. $T_g$-Depth Dependence of Nano-scale Films

Figure 5 shows the variation of o-Ps lifetime (top) and intensity (bottom) as a function of temperature for an 80 nm PS film at three different depths: near the surface (5 nm), the

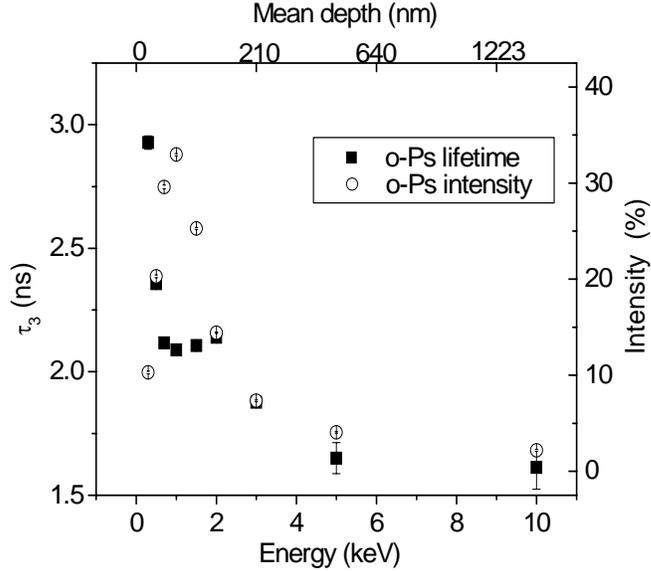

**Fig. 4** o-Ps lifetime ($\tau_3$) and intensity ($I_3$) vs. positron energy or mean depth (top) and free-volume radius distribution (bottom) in an 80-nm PS film. [7].

center of the film (36 nm), and near the interface (70 nm) with Si. In all plots, we observe a large expansion of $\tau_3$ at $T_g$ as due to the existence of free volume expansion form the glassy state to the rubbery state. The clear onset temperature of the fitted lines on $\tau_3$ represents $T_g$. For a thick polystyrene film, $T_g$ was determined to be $97 \pm 2$ °C, which is close to the DSC measurement ($100 \pm 1$ °C). For the thin film, we observe a significant suppression of $T_g$ near the surface ($80 \pm 5$ °C) and in the interface ($86 \pm 2$ °C), but no suppression at the center of the film ($98 \pm 3$ °C). The current result of lower $T_g$ near the surface and interface is consistent with the reported $T_g$ suppression near the surface in a thick film [7]. Furthermore, we only observe a slight trend of increase in $I_3$ across $T_g$. This is typical for polymeric materials where $I_3$ responds to temperature across $T_g$ less dramatic as o-Ps lifetime [7].

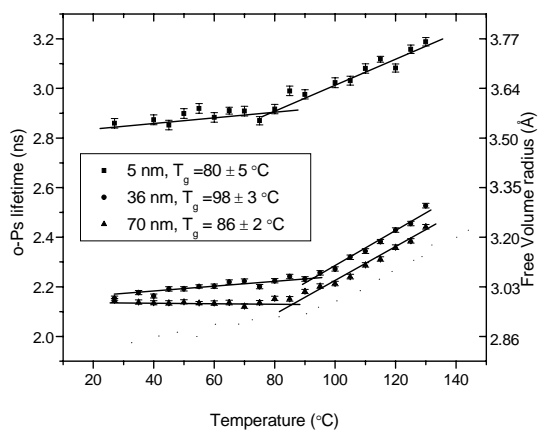

**Fig. 5** o-Ps lifetime (top) vs. temperature in an 80-nm polystyrene film at different depths and o-Ps lifetime (free-volume radius). The intercept of two fitted lines defines the $T_g$ [7].

We further analyzed all PAL data as a function of temperature into lifetime distribution for three depths of 80 nm PS film. In Figure 6 , we plotted the result of o-Ps lifetime (free-volume radius) distributions at different temperatures at the center (36 nm depth) of the PS film.

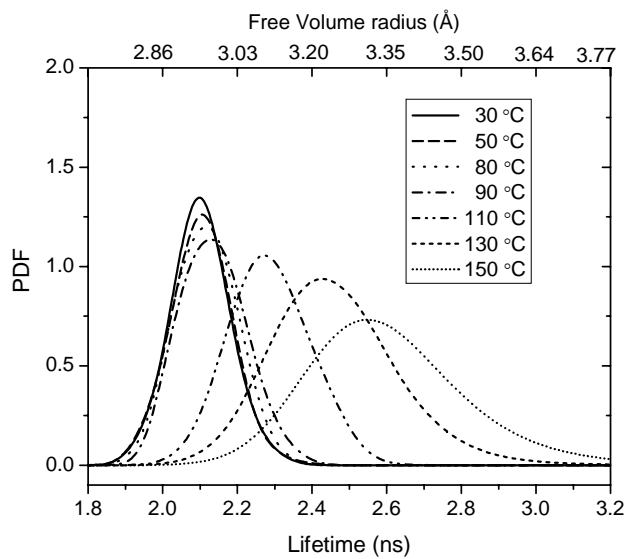

**Fig. 6** The o-Ps lifetime (radius) distribution functions at different temperatures at the center (36 nm) of an 80-nm polystyrene film. The bulk data from the thick PS film is included for comparison.

It is seen that the free-volume radius is distributed wider as the temperature increases and a large increase occurs at T. The measured FWHMs of the free-volume distributions at different depths are plotted vs. temperature in Figure 7. It is interesting to observe that the FWHM of the free-volume distribution has an onset temperature similar to the free-volume radius plots (Figure 5) and with a better precision (±1 C than that from hole size result in the determination of $T_g$. A broad free-volume distribution near the surface and at the interface indicates the existence of more end chains and the increase of polymeric chain mobility as evidence from the observed lower $T_g$. This result is similar to the depth dependence of polyurethane that the ortho-Ps lifetime is distributed wider as the positron incident energy is smaller, i.e. the free-volume distribution becomes wider as the mean depth is closer to the surface [11]. The broadening in free-volume distribution is also related to the degree of incomplete entanglement of polymer chains and to the increased chain motion near the surface and in the interface. For the currently studied polystyrene $M_w = 212,400$, the gyration radius is 13 nm calculated from $R_g(\text{Å}) = 0.28\ M_w{}^{1/2}$. It has a significant effect on polymer chain ends near the surface (5 nm) and the interface (70 nm) as seen in a broad free-volume distribution. This results a lower surface $T_g$ than that in the center of the thin film.

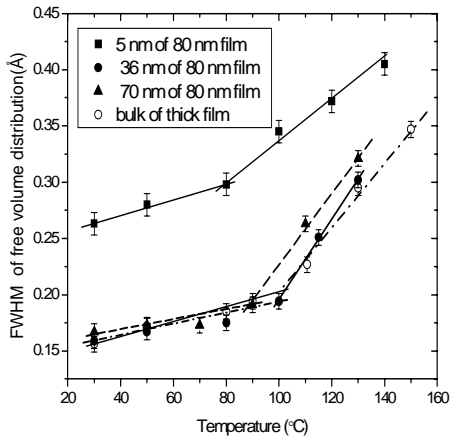

**Fig. 7** The FWHM of free-volume radius distributions o-Ps lifetime vs. temperature at different depth in an 80-nm polystyrene film. Lines were drawn through data points below and above $T_g$ [7].

The existing reported $T_g$ suppression for a thin film has been interpreted mainly in terms of incomplete entanglement of polymeric chains and broadening of $T_g$ [11]. The new information from the current PAL is the depth dependence of free-volume distributions, which are used to interpret the observed $T_g$ suppression near the surface and in the interface. For the interfacial $T_g$, since PS-Si is a weak interaction interface, we expect $T_g$ suppression (11 K) to be less than on the surface (17 K) where there is no interaction.

The current result of free-volume distribution by PAL offers a new interpretation that $T_g$ suppression in a thin film is due to different degrees of free-volume distribution at different depths. The free volume has the widest distribution near the surface which leads to large $T_g$ suppression; in the interface, it is slightly broadened but to a lesser extent, which leads to less $T_g$ suppression. In the center of the thin film, it has a distribution similar to that of thick film, which has no observable $T_g$ suppression in 80 nm polystyrene.

## 3. Sensitivity of Positron Annihilation with Cancerous Cells

We have performed two series of PAS experiments in search for sensitivity of positron annihilation with cancerous cells: the bulk skin samples with and without BCC and SCC samples using conventional PAL method [13], and depth profile studies of skin samples with and without BCC, SCC, and melanoma cells using the slow positron beam, PAL, and DBES methods [8].

Figure 8 shows the raw PAL spectra of the skin samples without (normal skin) and with cancer of squamous cell carcinoma (SCC). In the raw spectra of Fig.8 we can see that the skin sample without cancer cells has a larger Ps fraction than the skin sample with cancer cells. The PAL o-Ps results are analyzed and listed in Table 1.

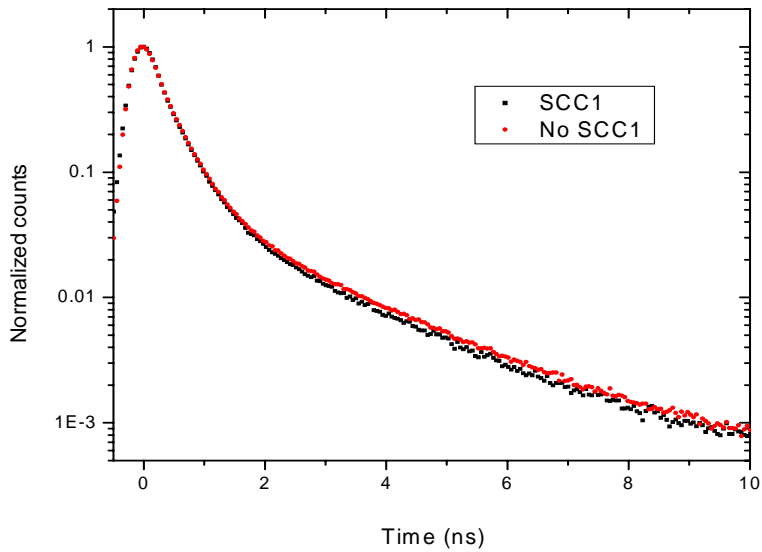

**Fig. 8** PAL raw spectra for the skin sample without (normal skin) and with cancer of squamous cell carcinoma (SCC) [13].

From this table we observe that in human skin the o-Ps intensity ($I_3$) is about 13-15% and the o-Ps lifetime ($\tau_3$) is between 1.9 to 2.0 ns. These data are similar to polymeric materials [3]. Also Comparing the values of $I_3$ and $\tau_3$ between the sample pair with and without cancer from the same patient we find both o-Ps $I_3$ and $\tau_3$ in this experiment under wet condition are consistent with the slow positron beam data at the skin depth > 5 keV (or 1 μm) [8].

**Table 1.** PAL analysis results for the skin samples without (normal) and with cancers of SCC (squamous cell carcinoma) and BCC

| | SCC #1 | No SCC #1 | BCC #2 | No BCC #2 |
|---|---|---|---|---|
| $\tau_3$ (ns) | 1.951 ± 0.011 | 2.043 ± 0.008 | 2.026 ± 0.005 | 2.081 ± 0.008 |
| $I_3$ (%) | 13.70 ± 0.14 | 14.65 ± 0.10 | 13.59 ± 0.06 | 14.16 ± 0.10 |
| R (Å) | 2.803 ± 0.008 | 2.887 ± 0.006 | 2.872 ± 0.004 | 2.921 ± 0.006 |
| fv (Å³) | 92.28 ± 0.83 | 100.8 ± 0.60 | 99.26 ± 0.42 | 104.4 ± 0.60 |
| ffv (%) | 2.275 ± 0.043 | 2.658 ± 0.035 | 2.428 ± 0.022 | 2.662 ± 0.035 |
| FWHM (Å) | 0.169 ± 0.009 | 0.252 ± 0.009 | 0.157 ± 0.009 | 0.190 ± 0.008 |

It is interesting to observe that the positron lifetimes in the skin with cancer of SCC and BCC are shorter than that in the normal skin. Since o-Ps lifetime ($\tau_3$) has a direct correlation with defect size, a smaller $\tau_3$ value in cancer samples indicates the reduction of free-volume size for the skin with cancer cells. As studied in polymeric materials, a reduction of free volume leads to the loss of durability of polymeric materials [5,6,10]. The intermolecular spaces inside cells are the possible free volume sites which can be considered as Ps localization and annihilation in the skin samples. Under this assumption, we estimate that the skin with cancer has a reduced free-volume size on the order of 10% as shown in Table 1 for the fv data. While the concentration of cancer cells in our sample is not known exactly, the detectable cancer cell from dermatological procedure is about 10-30% of the skin.

Similarly, we observe a difference of o-Ps intensity ($I_3$) between the samples with and without SCC and BCC cancer cells. We interpret this observation by two ways. First, comparing with normal skin, the positron has less affinity with the electrons from cancer cells in the skin to form Ps. Second, the secondary electrons in the positron radiation track due to the presence of strong electron scavengers inhibit the positron from forming Ps. The first interpretation may be similar to the reduction of o-Ps intensity due to micelle formation in surfactants[9]. For the second interpretation, a likely candidate for electron scavengers is free radicals inside cancer cells. Moreover, $I_3$ combining with free-volume size (from $\tau_3$) can be used to estimate the relative fractional free volume (ffv), which is a composite term to describe the number of holes and the size of free-volume. Here we find that the relative ffv values are reduced about 10% for both BCC and SCC cancer skin samples compared to those for normal skin samples. These results are consistent with that observed in the slow positron beam for the dried samples of BCC and SCC cancer skin as described below.

Furthermore, all PAL spectra were analyzed into continuous lifetime distributions (or radius distributions), as shown in Figure 8. Comparing with the corresponding distribution function of the skin sample with cancer, the distribution functions of the skin sample with cancer are found to be narrower than the normal skin. FWHMs of the distribution in the free-volume radius are listed in Table 1. Since the free-volume distribution is related to the degree of entanglement of macromolecular chains in the skin, a narrow distribution in the sample with cancer cells may indicate a less flexibility of molecular chain motion.

Next for the slow positron beam studies, we also observe a large difference in o-Ps intensity $I_3$ (%) between the samples with and without SCC or BCC cancer as shown in Figure 9.

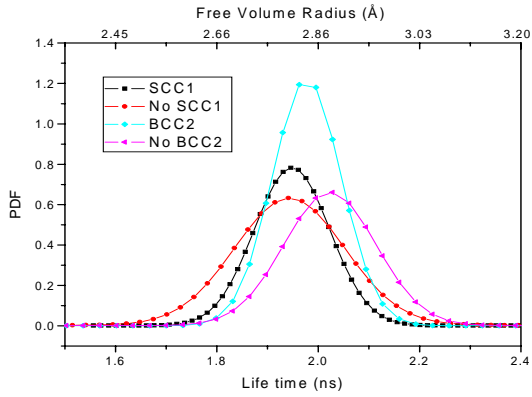

**Fig. 8** The o-Ps lifetime and free-volume hole radius distribution function for the skin samples without (normal) and with cancers of SCC (squamous cell carcinoma) and BCC (basal cell carcinoma) [13].

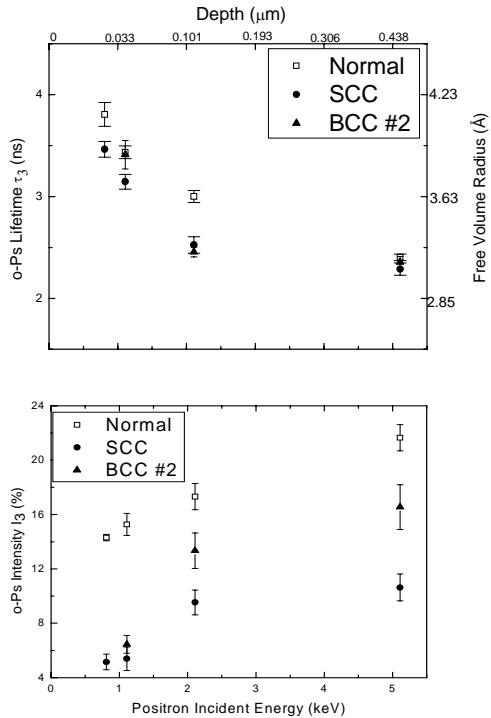

**Fig. 9** The o-Ps lifetime $\tau_3$ (free-volume radius ) and o-Ps intensity ($I_3$) vs. the positron incident energy (mean depth) in skin samples without (normal) and with Squamous Cell Carcinoma (SCC) or Basal Cell Carcinoma (BCC#2) [8]. [12].

The probability of o-Ps formation in the skin samples with cancer cells is found to decrease significantly to about one-third of its value in the normal skin near the surface and to about one-half in the deep skin. The interpretation of $I_3$ is not as straightforward as $\tau_3$ in terms of free volumes in molecular systems since it also depends on the radiation chemistry of the substances under study. Since the magnitude of the difference in $I_3$ is very significant, we interpret it in terms of both chemical and physical effects due to the presence of cancer cells. Chemically, a reduction of $I_3$ is often caused by a chemical inhibition to form Ps between the positron and available electrons or secondary electrons generated in radiation tracks of the positron [3]. Although the size of a cancer cell can vary (a basal cell carcinoma has an approximate mean diameter of 8.5 μm, and a squamous cell carcinoma has a mean diameter of 21.5 μm in the currently studied samples), it is still very large compared to the free volume (0.1-2 nm), and to the radiation track, which is also on the order of nm. Because of this vast size difference, the chemical effect should be interpreted at the molecular level. In the currently studied cancer skins, the fraction of cancer cells is estimated to be about 10-50% of the overall skin constituents. However, we observe a 100-200% reduction in $I_3$ in the cancer samples compared with that in the normal skin. This implies that the formation of o-Ps is selectively reduced 4 to 10-fold due to the presence of cancer cells in the skin. We have two theories for this observation: 1) The Ps and/or positron has an affinity toward reacting with chemical species inside cancer cells of the skin and/or 2) electrons in the radiation track are scavenged from Ps formation. A possible candidate for reacting species and/or electron scavengers is free radicals or reactive molecules inside cancer cells.

Furthermore, a change of both o-Ps lifetime ($\tau_3$) and intensity ($I_3$) indicates a change of free-volume fraction (ffv) in polymeric systems [2]. We estimated the relative ffv from free-volume hole radius (calculated from $\tau_3$) and o-Ps intensity ($I_3$) from an empirical formula [2]. It is interesting to observe a large decrease of ffv in the skin samples with cancerous cells compared with the sample without cancer. The reduction of ffv is 200-300%, particularly near the surface of the samples. We interpret the result as that cancer cells have a lower value of free volume than the normal cells. This infers that the cancer cells have a slow molecular mobility among molecular chains than the normal cell in the skin tissues.

The significant decrease of Ps formation probability ($I_3$), free-volume size ($\tau_3$), and relative free volumes (ffv) due to the presence of cancer cells in the skin is further confirmed by the S-parameter (a qualitative measure related to free volumes) data from DBES experiments as shown in Figure 9. In this figure, the S parameters in the samples, either with BCC or SCC, are significantly less than those in the skin without cancer as a function of depth from the surface to the bulk. The variation of S parameter vs. depth in skin samples is similar to results in polymeric coating systems [5,6,10]. A small and sharp rise of S near the surface is due to the back-diffusion and scattering of positrons and Ps into the vacuum. After the maximum, a decreasing S inside the surface between 2-7 μm is due to the variation in the chemical composition of the samples as detected by positron and Ps annihilation. A smaller S may be due to the contribution of inorganic elements, such as Na, K, Zn, Fe, etc. in the skin samples. The variation of S vs. depth can be fitted well in a model based on two layers (organic-rich surface and organic-inorganic layers) as in coatings [5,6], using a computer program VEPFIT [11].

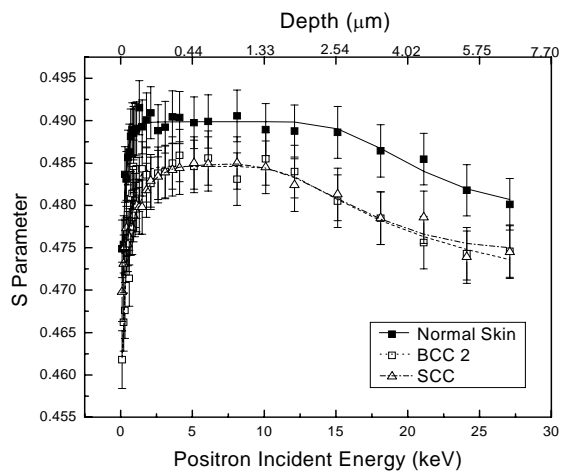

**Fig. 10** S parameter vs. positron incident energy (mean depth) in skin samples without (normal skin) and with cancer, Squamous Cell Carcinoma (SCC) or Basal Cell Carcinoma (BCC#2). Lines are fitted to a two-layer model using the VEPFIT program [8]

It is noted that the variation of S with respect to the depth is similar to that of $I_3$ of Figure 8 but at a smaller magnitude. This is because that S is primarily determined by p-Ps formation probability, which is one third of o-Ps formation probability as determined by $I_3$ [8]. In the same way, S is secondarily determined by the free-volume hole size as $\tau_3$ does of Figure 9 but it based on the uncertainty principle due the localization of Ps in free-volume holes. The overall result leads to that the S parameter in macromolecular systems is a quantity proportional to free volume. Although the S parameter is less quantitative for free volume determination than $\tau_3$, the data acquisition time for a DBES spectrum is one order of magnitude less than that for a PAL spectrum, which is beneficial for a shorter time required to develop PAS as a cancer detection and diagnosis tool in the future.

**Conclusion**

PAS is a novel spectroscopic method, which contains the most fundamental information about physical properties in molecules, i.e. wave function, electron density, and energy level. PAL and DBES techniques provide unique scientific information about the defect properties at the atomic and molecular levels. In this paper we demonstrate PAS's novelty and sensitivity in measuring the nanoscale layer structures in thin polymeric films and in detecting a difference in positron annihilation in the skin with and without cancer.

We have observed a significant variation of $T_g$ suppression as a function of depth in an 80 nm polystyrene thin film on Si: 17 K lower near the surface, and 11 K lower in the interface of the Si substrate than the center of the film or in the bulk. This depth dependence of $T_g$ suppression is interpreted as a broadening of free-volume distribution in the surface and interfaces. Although the positron has been discovered over 70 years, the uses and applications of PAS to molecular and biological systems are still at the developing stage and more systematic works along this line of research are needed in the near future.

PAL analysis shows human bulk skin samples with basal or squamous cell cancers under ambient condition reduce both $I_3$ and $\tau_3$ compared to those in normal skin. These results are consistent with those performed in the slow beam experiments of both PAL and DBES results under dried and high vacuum condition. The reduction of free volume contents indicates the loss of molecular flexibility after cancer formation. More experimental data in samples with and without different types of cancer are needed for further development of positron annihilation spectroscopy as a new noninvasive and external method for dermatology clinics, early detection of cancer, and nano-Positron Emission Tomography (PET) technology.

**Acknowledgments**

This research has been supported by NSF and NIST. We appreciate Drs. R. Suzuki and T. Ohdaira of AIST, X. Gu, and T. N, Nguyen of NIST for their collaboration. One of us (YCJ) would like to thank the partial support from the Ministry of Education, Taiwan and Prof. J.I. Lai for his sabbatical leave at Chung Yuan Christian University in Taiwan in 2007.